\documentstyle[12pt,emlines2]{article}
\newcommand{\tr}[1]{\,{\rm tr}\,#1\,}
\renewcommand{\theequation}{\thesection.\arabic{equation}}
\begin{document}
\title{
\begin{flushright}
{\small SMI-19-96 }
\end{flushright}
\vspace{2cm}
Composite p-branes in diverse dimensions}
\author{
Irina Aref'eva
\thanks
{ e-mail: arefeva@arevol.mian.su}
\\
{\it Steklov Mathematical Institute,}\\
{\it Vavilov 42, GSP-1, 117966, Moscow, Russia}\\
and \\
 Anastasia Volovich
 \thanks{e-mail address:
 nastya@arevol.mian.su}
  \\
{\it Department of Physics,$~$ Moscow State
 University}\\
{\it Vorobjevi gori, Moscow, 119899}\\
{\it and }\\
{\it   Mathematical College, Independent
  University of Moscow,}\\
 {\it 121108,PO BOX 68, Moscow,} {\it Russia }
}
\date {$~$}
\maketitle
\begin {abstract}
We use a simple  algebraic method to find a special class of composite
 p-brane solutions of higher dimensional gravity coupled with matter.
These solutions are composed of $n$ constituent p-branes corresponding
$n$ independent harmonic functions. A simple algebraic criteria of
existence of such solutions is presented. Relations with $D=11$, $D=10$
known solutions are discussed.
\end {abstract}

\newpage
\section{Introduction}
\setcounter{equation}{0}

The recent successful microscopic interpretation
of the Bekenstein-Hawking entropy within string theory \cite{SV}
has stimulated an investigation
of p-brane solutions \cite{DGHR}-\cite{gibbons}  of the low energy
field equations of the
string theory.  In view of suggestions \cite{M} that  D=11 supergravity
may be a low-energy effective field theory of a fundamental
"M-theory" which generalizes known string theories it is important to
find all classical p-brane solutions. It is also
interesting to understand what types of D=11 solutions may exist
in various dimension space-time.
Some p-brane solutions of supergravity theory may be associated with
non-perturbative state in superstring theory, like
monopoles or dyones in Yang-Mills theory.    
P-brane solutions of low-energy string action provide  also
an evidence for existence of string duality.

The simplest theories that are relevant
to superstring and supergravity theories are described by
actions

\begin{equation}
 I=\frac{1}{2\kappa ^{2}}\int d^{D}x\sqrt{-g}
 (R- \frac{1}{2(d+1)!}F^{2}_{d+1}),
                                           \label{I}
 \end{equation}

and

\begin{equation}
 I=\frac{1}{2\kappa ^{2}}\int d^{D}x\sqrt{-g}
 (R-\frac{(\nabla \phi)^2}{2} -\frac{e^{-\alpha \phi}}{2(d+1)!}F^{2}_{d+1}),
                                           \label{I'}
 \end{equation}
where $F_{d+1}$ is the $d+1$ differential form,
 $F_{d+1}=d{\cal A}_{d}$, $\phi$ is a dilaton.
In general, the fields $F$ and $\phi$ are  linear combinations of the
large variety of gauge fields and scalars.

The aim of this paper is to find  solutions of equations of motions
of the  theories (\ref{I}) and (\ref{I'}) that have so-called thin structure.
In the p-brane terminology  \cite{LPS,PapTow1,Ts1,Pap,Ts2}
these solutions correspond to $n$ constituent  p-branes.
We will use an algebraic
methods (cf.\cite {Duff,AV,AVV}) of finding  solutions.
The advantage of this simple method is its universality.
The cases of different dimensions and different
set of matter fields are treated in the same way.
According to
this method one can reduce the problem of finding solutions to
an algebraic problem of solving an overdetermined non-linear system of
algebraic equations. To make this reduction one uses algebraic requirements
(see eqs.(\ref{27}) below) which provide an absence of forces
acting on matter sources and permit to reduce equations for matter
fields to the harmonically conditions of functions describing matter
fields. These conditions are direct analogues of "no-force"
condition which has been recently used
by Tseytlin \cite{Ts212} to get various BPS
configurations in $D=11$ and $D=10$ cases
(see also\cite{DGHR,duff1,Cal,Duff}). These results
are consistent with other approaches (D-brane supersymmetry
analysis or study of potential between D-branes 
\cite{pol1,Gr,Lip,PapTow1,Ts1,kl,Gaun}).
We will consider arbitrary dimension  space-time  and the case of
"electric" solutions for one antisymmetric field.
We will find an algebraic criteria of existence of solutions
with thin structure. This criteria gives a restriction on
dimension of  space-time
and parameters $d$  and $\alpha $.
For particular cases $D=10$  and $D=11$ this criteria
is satisfied and  some solutions
reproduce the solutions recently  found  by Tseytlin \cite{Ts2}
and Papadopoulos and Townsend \cite{PapTow2} that  after reductions
produce supersymmetric BPS saturated p-brane solutions of low-dimensional
theories \cite{duff1,Gu,PapTow1}.
A generalization to the case of magnetic solutions is straightforward,
however a generalization for more fields with different dimensions
need more treatment.This consideration is important in the context
of \cite{Berg}-\cite{Lust}.

\section{Gravity $+$ d-form}

We will use the following ansatz for the metric
\begin{equation}
ds^{2}=e^{2A}\eta_{\mu \nu} dy^{\mu}
dy^{\nu}+\sum _{i=1}^{n}e^{2F_{i}}dz_{i}^{m_{i}}dz_{i}^{m_{i}}
+e^{2B}dx^{\gamma}dx^{\gamma},
 \label{1}
\end{equation}
$\mu$, $\nu$ run from $0$ to $q-1$, $\eta_{\mu\nu}$
is a flat Minkowski metric, $m_{i},n_{i}$ run from $1$ to $r$,
$\alpha,\beta,\gamma$ run from
$1$ to $\tilde{d}+2$.
$A$, $B$
 and $C$ are functions on $x$.
Under an assumption
\begin{equation}
qA+\sum r_{i}F_{i} +{\tilde d}B=0,
                                        \label{9}
\end{equation}
the Ricci tensor for metric (\ref{1}) has a simple  form
 \begin{equation}
 R_{\mu\nu}=-h_{\mu\nu}e^{2(A-B)}\Delta  A,~~
~~ R_{m_{i}n_{i}}=-\delta_{m_{i}n_{i}}e^{2(F_{i}-B)}\Delta  F_{i}
 \label{10}
\end{equation}
\begin{equation}
R_{\alpha\beta}=- q\partial_{\alpha} A\partial_{\beta} A -
\sum _{i} r_{i}\partial_{\alpha} F_{i}\partial_{\beta} F_{i}
+
{\tilde d}\partial_{\alpha} B\partial_{\beta} B
-\delta_{\alpha\beta}\Delta B
                                         \label{12}
\end{equation}

For matter field  we use ansatz
\begin{figure}
\begin{center}
\special{em:linewidth 0.4pt}
\unitlength 0.30mm
\linethickness{0.4pt}


\end{center}
\end{figure}
\begin{equation}
{\cal A} =dy^{0}\wedge dy^{1}\wedge ...\wedge dy^{q-1}\wedge[
dz^{1}_{1}\wedge...\wedge dz^{r_{1}}_{1}h_{1}e^{C_{1}}+
...+dz^{1}_{n}\wedge...\wedge dz^{r_{n}}_{n}h_{n}e^{C_{n}}],
                                                       \label{21}
\end{equation}

here $r_{i}=r$, $i=1,..n$, i.e. $~D=q+nr+\tilde d+2$, and for $r \neq 
1$ the components of energy-momentum tensor read \begin{equation} 
T_{\mu \nu}=-\eta_{\mu \nu} e^{2(A-B)}\sum _{i=1}^{n}
\frac{h_{1}^{2}}{4}e^{-2qA-2rF_{i}
-2C_{i}}(\partial C_{i})^{2}
                                              \label{22}
\end{equation}
\begin{equation}
T_{m_{i}n_{i}}=\delta_{m_{i}n_{i}} e^{2(F_{i}-B)}
[-\frac{h_{i}^{2}}{4}e^{-2qA-2rF_{i}+2C_{i}}(\partial C_{i})^{2}
+\sum _{j\neq i}^{n}\frac{h_{j}^{2}}{4}e^{-2qA-2rF_{j}+2C_{j}}
(\partial C_{j})^{2}]
                                             \label{23}
\end{equation}
\begin{equation}
T_{\alpha \beta}=
-\sum _{i=1}^{n}\frac{h_{i}^{2}}{2}e^{-2qA-2rF_{i}+2C_{i}}
[\partial_{\alpha} C_{i}\partial_{\beta} C_{i}
-\frac{\delta_{\alpha \beta} }{2}
(\partial C_{i})^{2}]
                                             \label{26}
\end{equation}
 If we assume that
\begin{equation}
 qA+rF_{i}=C_{i},~~i=1,...n
                                              \label{27}
\end{equation}
the form of $T_{MN}$ crucially  simplifies
and the Einstein equations have the form
\begin{equation}
 \Delta  A=\sum _{i}th_{i}^{2}(\partial C_{i})^{2},
                                               \label{33}
\end{equation}

\begin{equation}
\Delta  F_{i}=th_{i}^{2} (\partial C_{i})^{2}
-\sum _{j\neq i}uh_{j}^{2}(\partial C_{j})^{2}
                                           \label{34}
\end{equation}

$$
- q\partial_{\alpha} A\partial_{\beta} A -
\sum _{i} r_{i}\partial_{\alpha} F_{i}\partial_{\beta} F_{i}
-{\tilde d}\partial_{\alpha} B\partial_{\beta} B
- \delta_{\alpha\beta}\Delta B=
$$
\begin{equation}
-\sum _{i=1}^{n}[\frac{h_{i}^{2}}{2}
\partial_{\alpha} C_{i}\partial_{\beta} C_{i}-u\delta_{\alpha \beta}]
(\partial C_{i})^{2}]
                                         \label{35}
\end{equation}
where $t$ and $u$ are given by
\begin{equation}
                                      \label{31}
t=\frac{1}{2}\cdot\frac{D-2-q-r}{D-2},~~
u=\frac{1}{2}\cdot\frac{q+r}{D-2}
 \end{equation}

The equation of motion for the antisymmetric field,
\begin{equation}
\partial _{M}(\sqrt{-g}F^{MM_{1}...M_{d}})=0,
                                           \label{16}
\end{equation}
 under conditions
(\ref{9}) and (\ref{27})
for the ansatz (\ref{21}) reduces to
\begin{equation}
\partial _{\alpha}(e^{ -C_{i}}\partial _{\alpha}C_{i})=0,~
\mbox{or}~ \Delta C=(\partial C)^{2}.
                                         \label{36}
\end{equation}
The form of equation (\ref{36}) shows a physical meaning of conditions
(\ref{27}). Under these conditions equations of motion for
the antisymmetric field reduce to the harmonicity conditions
for functions $\exp (-C_{i})$  describing matter field, i.e.
there are no sources for $\exp (-C_{i})$.
These conditions correspond to "no-force" condition of vanishing of static
force   on a q-brane probe in gravitational background produced by another
p-brane \cite{DGHR,duff1,Cal,Duff,Ts212}.
The LHS of (\ref{27}) is the logarithm of the "Nambu' term for $q+r$-brane
probe in the gravitational background (\ref{1}) and the RHS of (\ref{27})
is the logarithm of the "Wess-Zumino" term.

We solve equations (\ref{33})and  (\ref{34})  assuming
\begin{equation}
 A=t\sum _{i}h_{i}^{2} C_{i},
                                               \label{37}
\end{equation}

\begin{equation}
F_{i}=th_{i}^{2} C_{i}
-u\sum _{j\neq i}h_{j}^{2} C_{j}
                                      \label{38}
\end{equation}

To cancel the terms in front of $\delta_{\alpha \beta}$ in equation
(\ref{35}) we also assume
\begin{equation}
B=-u\sum _{i} h_{i}^{2} C_{i}
                                         \label{39}
\end{equation}

Note that  if we want  to solve equations  for arbitrary
harmonic functions $H_{i}=$ $\exp (-C_{i})$ we have also to assume the
conditions which follow  from (\ref{27}), (\ref{9}) and nondiagonal
part of (\ref{35}). Equations  (\ref{27}) get the relations
\begin{equation}
                                \label{40.0}
(q+r)t h_{i}^{2} C_{i}+
(qt-ru)\sum _{j\neq i} h_{j}^{2} C_{j}=C_{i},~~~i=1,...n
\end{equation}
which under the assumption of independence of $C_{i}$
give
\begin{equation}
qt-ru=0,                                      \label{40}
\end{equation}

\begin{equation}
(q+r)th_{i}^{2} =1.
                                      \label{41}
\end{equation}
Since $t$ and $u$ are given by the formulae (\ref{31})
the condition (\ref{40}) makes a restriction on dimensions $D$, $q$
and $r$

\begin{equation}
q(D-2)=(q+r)^{2}
                                         \label{42}
\end{equation}
Note that under this assumption the formula (\ref{31})
 has the form
\begin{equation}
t=\frac{1}{2}\frac{r}{q+r},~~u=\frac{1}{2}\frac{q}{q+r}
                                         \label{31'}
\end{equation}

Equation (\ref{41}) gives

\begin{equation}
h_{i}^{2} =h^{2}\equiv \frac{1}{(q+r)t},
                                      \label{43}
\end{equation}

Equation (\ref{9}) gives
\begin{equation}
qt+r(t-(n-1)u)-\tilde{d}u=0,
                                      \label{44}
\end{equation}
Since $D=q+rn+\tilde{d}+2$ and $qt=ru$ equation (\ref{40})
is equivalent to $q(D-2)=(q+r)^{2}$, that we have just assumed.

Equation (\ref{35}) gives  two types of relations
(\ref{39}). One type of relations
follows from the requirement of compensation of terms
$\partial_{\alpha} C_{i}\partial_{\beta} C_{i}$
in the both sides of equation (\ref{35}) and the second one
produces a compensation of mixed terms
$\partial_{\alpha} C_{i}\partial_{\beta} C_{j}$,$i\neq j$.
Straightforward calculations show that the both type of terms compensate.

This calculation shows that the form of metric is
\begin{equation}
ds^{2}=(H_{1}H_{2}...H_{n})^{-4t/r}\eta_{\mu \nu} dy^{\mu}
dy^{\nu}+
(H_{1}H_{2}...H_{n})^{4u/r)}[
\sum _{i=1}^{n}H_{i}^{-2/r}dz_{i}^{m_{i}}dz_{i}^{m_{i}}
+dx^{\gamma}dx^{\gamma}],
 \label{50.0}
\end{equation}
or due to (\ref{31'})

$$ds^{2}=(H_{1}H_{2}...H_{n})^{2q/r(q+r)}[
(H_{1}H_{2}...H_{n})^{-2/r}\eta_{\mu \nu} dy^{\mu}
dy^{\nu}+$$

\begin{equation}
(H_{1})^{-2/r}dz_{1}^{m_{1}}dz_{1}^{m_{1}}+...
+(H_{n})^{-2/r}dz_{n}^{m_{n}}dz_{n}^{m_{n}}+
dx^{\gamma}dx^{\gamma}],
 \label{51}
\end{equation}

In Section 4 we will present solutions of  equation (\ref{42}).

\section{Gravity $+$ dilaton  $+$  d-form}

We use the same ansatz for the metric and for
the antisymmetric field but now in addition to the relation (\ref{9})
we suppose the following conditions
\begin{equation}
 -\alpha \phi -2q^{(\alpha)}A^{(\alpha)}-2r^{(\alpha)}F^{(\alpha)}_{i}
 -2C_{i}=0,~i=1,...n
                                              \label{2.27}
\end{equation}
In this case the form of $\mu \nu$ and $m_{i}n_{i}$
components of the Einstein
equations does not change
\begin{equation}
 \Delta  A^{(\alpha)}=\sum _{i}t^{(\alpha)}
 h^{(\alpha)2}_{i}(\partial C_{i})^{2},
                                               \label{2.33}
\end{equation}

\begin{equation}
\Delta  F^{(\alpha)}_{i}=t^{(\alpha)}h^{(\alpha)2}_{i}
 (\partial C_{i})^{2}
-\sum _{j\neq i}u^{(\alpha)}h^{(\alpha)2}_{j}(\partial C_{j})^{2}
                                           \label{2.34}
\end{equation}
and there is one extra term in the RHS of $\alpha \beta$ equations
$$
- q\partial_{\alpha} A\partial_{\beta} A -
\sum _{i} r_{i}\partial_{\alpha} F_{i}\partial_{\beta} F_{i}
-{\tilde d}\partial_{\alpha} B\partial_{\beta} B
- \delta_{\alpha\beta}\Delta B=
$$
\begin{equation}
-\sum _{i=1}^{n}[\frac{h^{(\alpha)2}_{i}}{2}
\partial_{\alpha} C_{i}\partial_{\beta} C_{i}-u\delta_{\alpha \beta}]
(\partial C_{i})^{2}]
                                         \label{2.35}
\end{equation}
Now the parameters  $t^{(\alpha)}$ and $u^{(\alpha)}$
are given by the formula
\begin{equation}
t^{(\alpha)}=\frac{1}{2}[1-\frac{q^{(\alpha)}+r^{(\alpha)}}
{D^{(\alpha)}-2}],~~
                                          \label{2.31}
u^{(\alpha)}=\frac{1}{2}\frac{q^{(\alpha)}+r^{(\alpha)}}
{D^{(\alpha)}-2}]
                                       \label{2.32}
\end{equation}
Equations of motion (\ref{16}) under conditions
(\ref{9}) and (\ref{2.27})
for the ansatz (\ref{21}) reduce as in the previous case
to  the harmonicity conditions (\ref{2.36})
for functions $\exp (-C_{i})$.
Equation of motion for dilaton reads
\begin{equation}
\Delta \phi -\frac{\alpha}{2}\sum _{i}
 h^{(\alpha)2}_{i}(\partial C_{i})^{2}
                                         \label{2.36}
\end{equation}

We solve equation (\ref{2.36}) assuming
\begin{equation}
\phi=\frac{\alpha}{2}\sum _{i}h^{(\alpha)2}_{i} C_{i},
                                               \label{2.37}
\end{equation}
and we solve equations  (\ref{2.33}) and  (\ref{2.34}) the  relation
(\ref{2.37}) (now we put indexes $\alpha $
for coefficients).

To cancel the terms in front of $\delta_{\alpha \beta}$ in equation
(\ref{2.35}) we also assume
\begin{equation}
B^{(\alpha)}=-u^{(\alpha)}\sum _{i} h^{(\alpha)2}_{i} C_{i}
                                         \label{2.39}
\end{equation}

Note that  if we want  to solve equations  for arbitrary
harmonic functions $H_{i}=$ $\exp (-C_{i})$ we have also to assume the
conditions which follow  from (\ref{2.27}), (\ref{9}) and nondiagonal
part of (\ref{2.35}). Equations  (\ref{2.27}) get the relations
\begin{equation}
                                \label{2.40.0}
[\frac{\alpha ^{2}}{4}
+(q^{\alpha)}+r^{\alpha)})t^{(\alpha)}]h_{i}^{(\alpha)2} C_{i}+
[\frac{\alpha ^{2}}{4}+q^{\alpha)}t^{\alpha)}-
r^{(\alpha)}u^{(\alpha)}]\sum _{j\neq i} h_{j}^{(\alpha)2} C_{j}
=C_{i},~~~i=1,...n
\end{equation}
which under the assumption of independence of $C_{i}$
give

\begin{equation}
\frac{\alpha ^{2}}{4}=r^{(\alpha)}u^{(\alpha)}-
q^{(\alpha)}t^{(\alpha)},
                                      \label{2.40}
\end{equation}

\begin{equation}
[\frac{\alpha^{2}}{4}+(q^{(\alpha)}+r^{(\alpha)})t^{(\alpha)}]
h^{(\alpha)2}_{i} =1.
                                      \label{2.41}
\end{equation}
Since $t^{(\alpha)}$ and $u^{(\alpha)}$
are given by the formulae (\ref{2.31})
the condition (\ref{2.40}) makes a restriction on dimensions $D^{(\alpha)}$,
$q^{(\alpha)}$
and $r^{(\alpha)}$

\begin{equation}
(\frac{\alpha ^{2}}{2}+q^{(\alpha)})(D^{(\alpha)}-2)
=(q^{(\alpha)}+r^{(\alpha)})^{2}
                                         \label{2.42'}
\end{equation}
Note that under this assumption the formulae (\ref{2.31})
have the form
\begin{equation}
u^{(\alpha)}=\frac{1}{4}\frac{2q^{(\alpha)}+\alpha ^{2}}
{q^{(\alpha)}+r^{(\alpha)}},~~
t^{(\alpha)}=\frac{1}{4}\frac{2r^{(\alpha)}-\alpha ^{2}}
{q^{(\alpha)}+r^{(\alpha)}}
                                         \label{2.31'}
\end{equation}

The LHS of (\ref{2.41}) for $t^{(\alpha)}$  and $u^{(\alpha)}$
given by formula (\ref{2.31'})
can be represented as $r^{(\alpha)}h^{(\alpha)2}/2$, and , therefore,
equation (\ref{2.41}) gives

\begin{equation}
h^{(\alpha)2}_{i} =h^{(\alpha)2}\equiv \frac{2}{r^{(\alpha)}},
                                      \label{2.43}
\end{equation}

As in the case of absence of dilaton one can check that for
$\alpha$, $q^{(\alpha)}$, $r^{(\alpha)}$ and $D^{(\alpha)}$
satisfying relation (\ref{2.42'}) and $h^{(\alpha)}$ given by
(\ref{2.43}) equations (\ref{9}) as well as equations
given a compensation of terms
$\partial_{\alpha} C_{j}\partial_{\beta} C_{i}$ ($i=j$ as well as $i\neq
j$)in the both sides of equation (\ref{2.35}) are fulfilled.

This calculation shows that the  metric
$$
ds^{2}=(H_{1}H_{2}...H_{n})^{-4t^{(\alpha)}/r^{(\alpha)}}\eta_{\mu \nu} dy^{\mu}
dy^{\nu}+
$$
\begin{equation}
(H_{1}H_{2}...H_{n})^{4u^{(\alpha)}/r^{(\alpha)})}[
\sum _{i=1}^{n}H_{i}^{-2/r^{(\alpha)}}dz_{i}^{m_{i}}dz_{i}^{m_{i}}
+dx^{\gamma}dx^{\gamma}]=
 \label{2.50.0}
\end{equation}
$$
(H_{1}H_{2}...H_{n})^{4u^{(\alpha)}/r^{(\alpha)}}[
(H_{1}H_{2}...H_{n})^{-2/r^{(\alpha)}}
\eta_{\mu \nu} dy^{\mu}dy^{\nu}+   $$

$$
\sum _{i=1}^{n}H_{i}^{-2/r^{(\alpha)}}dz_{i}^{m_{i}}dz_{i}^{m_{i}}
+dx^{\gamma}dx^{\gamma}],$$

 and  matter fields in the form
\begin{equation}
\exp \phi=(H_{1}H_{2}...H_{n})^{-\alpha /r^{(\alpha)}}
 \label{2.51}
\end{equation}
\begin{equation}
{\cal A} =h^{(\alpha)}dy^{0}\wedge dy^{1}\wedge ...\wedge dy^{q-1}\wedge[
dz^{1}_{1}\wedge ...\wedge dz^{r}_{1}H^{-1}_{1}+...+
dz^{1}_{n}\wedge ...\wedge dz^{r}_{n}H^{-1}_{n}]
                                                       \label{21a}
\end{equation}
are the solution of the theory.

\section{Examples}
\setcounter{equation}{0}

Note that equation (\ref{42}) is very restrictive since it has to be solved
for integers.
Let us present some  examples.

For dimensions $D=4,5,6,7,8$  and $9$  there are solutions only
with $r=0$
and we have 2-block solutions with   $\tilde{d}=0$. In these cases, either the
spacetime is asymptotically $M^{q}\times Y$, where $Y$ is a two-dimensional
conical space, or the metric exhibits logarithmic behavior
as $|x|\to \infty$ \cite{Bgreen}-\cite{Lu}.

We get more interesting structures in $D=6$ case.
There are four types of solutions.

 $i)~q=1,~ r=1, n=2, \tilde{d}=1,$

 $ii)~q=1,~ r=1, n=3, \tilde{d}=0,$

 $iii)~q=1,~ r=1, n=4, \tilde{d}=-1,$

 $ iv)~~q=1,~ r=1, n=5, \tilde{d}=-2.$
\\Here we have to assume that different branches with
$r=1$ correspond to different gauge field
$${\cal A}^{(I)}=h dy^{0}\wedge dz_{i}H_{i}^{-1}\delta_{iI} ,
$$
I=1,...n.(Otherwise we cannot guarantee the diagonal form of the
stress-energy tensor).
The solution iv) is identified with the Minkowski
vacuum of the  theory.
The solution iii) separates the 6-dimensional
space-time into two asymptotic regions like a domain wall \cite{Lu}.
 The metric for the solution with $\tilde{d}=0$ has a logarithmic behavior,
$H=\sum _{a}ln
(\frac {Q_{a}}{|x-x_{a}|^{2}})$.

For $\tilde{d}=1$ we have
\begin{equation}
 \label{D6}
ds^{2}=(H_{1}H_{2})[-(H_{1}H_{2})^{-2}dy_{0}^{2}+(H_{1})^{-2}dz_{1}^{2}
                    +(H_{2})^{-2}dz_{2}^{2}+(dx_{i})^{2}] ,
                     i=1,2,3
\end{equation}
$${\cal A}^{(1)}=\sqrt{2}dy_{0}\wedge dz_{1}H_{1}^{-1},
$$

$${\cal A}^{(2)}=\sqrt{2}dy_{0}\wedge dz_{2}H_{2}^{-1},
$$
\begin{equation}
 \label{Har1}
  H_{1}=1+\sum_{a=1}^{l_{1}} \frac{Q_{a}^{(1)}}{|x-x_{a}^{(1)}|};~~
  H_{2}=1+\sum_{b=1}^{l_{2}} \frac{Q_{b}^{(2)}}{|x-x_{b}^{(2)}|},~~
|x-x_{a}|=(\sum _{i=1}^{3}|x_{_{i}}-x_{a_{i}}|^{2})^{1/2}.
 \end{equation}

If $l_{1}=l_{2}=l$ and $x_{a}^{(1)}=x_{a}^{(2)}=x_{a}$ the
metric (\ref{D6}) has horizons at the points $x_{a}$.
The area (per unit of length in all p-brane directions)
of the horizons $x=x_{a}$ is
\begin{equation}
 \label{H6}
A_{4}= 4\pi{ \sum_{a=1}^{l}}Q_{a}^{(1)} Q_{a}^{(2)}
\end{equation}
This confirms an observation {\cite {kallosh}} that extremal
black holes have non vanishing event horison in the presence of two
or more charges (electric or magnetic).There are more solutions for several
scalar fields \cite{Em}-\cite{Lust}.

For $D=10 $ we have two solutions with $\tilde {d}=0$.

 $~i)~~~ q=8,~ r=0, \tilde{d}=0, n=1$;

 $~ii)~ q=2,~ r=2,~ \tilde{d}=0,~n=3.$
\\There is also solution with
$q=2,~ r=2,~ \tilde{d}=2,~n=2,$

$$
ds^{2}=(H_{1}H_{2})^{1/2}[(H_{1}H_{2})^{-1}(-dy_{0}^{2}+dy_{1}^{2}+Kdu^{2})
$$
\begin{equation}
 \label{D10}
+H_{1}^{-1}(dz_{1}^{2}+dz_{2}^{2})+H_{2}^{-1}(dz_{3}^{2}+dz_{4}^{2})
+\sum _{i=1}^{4}(dx_{i})^{2}] ,
 \end{equation}
    $$   H_{1}=1+\sum_{a} \frac{Q_{a}^{(1)}}{|x-x_{a}|^{2}}; ~
    H_{2}=1+\sum_{b} \frac{Q_{b}^{(2)}}{|x-x_{b}|^{2}},$$

$$ K={\sum_{a}}\frac{Q_{a}}{|x-x_{a}|^{2}};~~~~~u=y_{0}+y_{1}$$
The area of this horizon (per unit length in all p-brane directions)
is
\begin{equation}
 \label{H10}
A_{8}= \omega _{3}\sum _{a} (Q_{a}^{(1)} Q_{b}^{(2)} Q_{a})^{1/2}
\end{equation}
where
$\omega_{3}=2\pi ^{2} $ is   the area  of the  unit 3-dimension sphere.
In this case the different components of the same gauge field act
as fields corresponding to different charges.
For $D=11$  we have the following solutions with $\tilde {d}>1$.\\
$i)~~q=1,~ r=2,~ \tilde{d}=2,~n=3,$
$$
ds^{2}=(H_{1}H_{2}H_{3})^{1/3}[-(H_{1}H_{2}H_{3})^{-1}dy_{0}^{2}
$$
\begin{equation}
 \label{D10'}
+H_{1}^{-1}(dz_{1}^{2}+dz_{2}^{2})+H_{2}^{-1}(dz_{3}^{2}+dz_{4}^{2})
+H_{3}^{-1}(dz_{5}^{2}+dz_{6}^{2})+{\sum_{i=1}^{4}}dx_{i}^{2}],
 \end{equation}
    $$   H_{c}=1+\sum_{a} \frac{Q_{a_{c}}}{|x-x_{a}|^{2}};
    ~~c=1,2,3. $$
For $H_{3}$ this solution reproduces  a solution found in \cite{Ts212}.
We get non-zero area of the horizon $x=x_{a}$.
\begin{equation}
 \label{H11}
A_{9}= \omega _{3}\sum _{a} (Q_{a}^{(1)} Q_{a}^{(2)}Q_{a}^{(3)})^{1/2}
\end{equation}
$~ii) ~q=4,~ r=2, \tilde{d}=1,~n=2,$
$$
ds^{2}=(H_{1}H_{2})^{2/3}[(H_{1}H_{2})^{-1}(-dy_{0}^{2}+dy_{1}^{2}
+dy_{2}^{2}+dy_{3}^{2})
$$
\begin{equation}
 \label{D11}
+H_{1}^{-1}(dz_{1}^{2}+dz_{2}^{2})+H_{2}^{-1}(dz_{3}^{2}+dz_{4}^{2})
+{\sum_{i=1}^{3}}dx_{i}^{2}],
\end{equation}
where   $  H_{1}$ and $H_{2}$ are given by
(\ref{Har1}). This solution has been recently found in
\cite{Ts212}.
The area of the horizon $x=x_{a}=x_{b}$ is equal to zero.

For $ D=12,13,14,15,16,17,$ there are only
$\tilde{d}=0 $  solutions.

For $D=18$ and $D=20$ there are  the following
solutions with $\tilde {d}>1$.

$D=18,~i)~~q=1,~ r=3, \tilde{d}=3,~n=4,~H_{c}=\sum
\frac {Q_{a_{c}}}{(x-x_{a_{c}})^{3}}.~ c=1,2,3,4.$

$~~~~~~~~~~~~ii)~q=4,~ r=4, \tilde{d}=4,~n=2,~H_{c}=\sum
\frac {Q_{a_{i}}}{(x-x_{a_{c}})^{4}}, c=1,2$

$ D=20,~ ~q=5,~ r=4, \tilde{d}=1,~n=3,~H_{c}=\sum
\frac {Q_{a_{c}}}{(x-x_{a_{c}})^{3}},~ c=1,2,3$

Let us present some  examples of solution of equation (\ref{2.31'}).
Note that this equation gives a "quantized" values for $\alpha$.

For $D=10$ we have  $q=1,~ r=3,$ $\alpha =\pm \sqrt{2}, \tilde{d}=1,$
$n=2$
and the corresponding metric has the form
$$
ds^{2}=(H_{1}H_{2})^{1/3}[-H_{1}H_{2})^{-2/3}dy_{0}^{2}+H_{1}^{2/3}(dz_{1}^{2}
+dz_{2}^{2}+dz_{3}^{2})
$$
\begin{equation}
 \label{D11'}
+H_{2}^{-2/3}(dz_{4}^{2}+dz_{5}^{2}+dz_{6}^{2})
+\sum_{i=1}^{3}dx_{i}^{2}],
\end{equation}
where   $  H_{1}$ and $H_{2}$ are given by
(\ref{Har1}). This solution correspond to IIA supergravity.
The area of the horizon $x=x_{a}=x_{b}$ is equal to zero.

For $ D=11$ we have the following solutions: $i)~q=1,~ r=5,
~\alpha =\pm \sqrt{6},~\tilde{d}=3,~n=1;$
 ii)$~q=2,~ r=4, ~\alpha =\pm 2,
~\tilde{d}=3,~n=1,$
iii)$~q=3,~ r=3, ~\alpha =\pm \sqrt{2},~\tilde{d}=3,~n=1.$

In conclusion, we have found multi-block p-brane solutions
for high dimensional gravity interacting with matter.
We have assumed the "electric" ansatz for matter field and using
"non-force"
conditions for local fields (\ref{27}) together with the
harmonic gauge condition
(\ref{9}) to reduce the system of differential equations
to overdetermined  system of non-linear equations.
The found solutions support
a picture in which
the extremal p-brane can be viewed as a composite of
`constituent' branes, each of the latter possessing a charge
corresponding to one of the gauge fields or only to one of
the comonents for the decomposition (Fig.1).
It would be  interesting to perform
the similar calculations for "dyonic" ansatz
as well as to consider reduction of found solutions to low dimensional
 cases a la  Kaluza-Klein   \cite{PapTow2}.

\section{Acknowledgments}
We are grateful to A.Tseytlin for useful comments.
This work is supported by the RFFI grant 96-01-00608.
We thank M.G.Ivanov for taking our attention to a special
case of $r_{i}=1$.

\end{document}